\documentclass[fleqn,twoside]{article}
\usepackage{espcrc2}

\usepackage{graphicx}
\usepackage[figuresright]{rotating}

\newcommand{\AmS}{{\protect\the\textfont2
  A\kern-.1667em\lower.5ex\hbox{M}\kern-.125emS}}

\title{Double-Chooz: a search for $\theta_{13}$}

\author{Th. Lasserre\address[MCSD]{DSM/DAPNIA/SPP, CEA/Saclay, 91191 Gif-sur-Yvette, France}}%
       
\begin{document}

\begin{abstract}
The Double-Chooz experiment goal is to search for a non-vanishing
value of the $\theta_{13}$ neutrino mixing angle. This is the last step to
accomplish prior moving towards a new era of precision measurements 
in the lepton sector. The current best constraint on the third mixing
angle comes from the CHOOZ reactor neutrino experiment
$\sin(2\theta_{13})^{2}<0.2$ (90$\%$ C.L., $\Delta m_{atm}^{2}=2.0$ eV$^{2}$).   
Double-Chooz will explore the range of $\sin(2\theta_{13})^{2}$
from 0.2 to 0.03-0.02, within three years of data taking. The improvement
of the CHOOZ result requires an increase in the statistics, a reduction
of the systematic error below one percent, and a careful control of
the backgrounds. Therefore, Double-Chooz will use two identical detectors,
one at 150 m and another at 1.05 km distance from the
Chooz nuclear cores. In addition, we will to use the near
detector as a ``state of the art'' prototype to investigate the
potential of neutrinos for monitoring the civil nuclear power plants. 
The plan is to start operation with two detectors in 2008, and to
reach a sensitivity sin$^{2}$$(2\theta_{13})$ of 0.05 in 2009, and
0.03-0.02 in 2011.
\vspace{1pc}
\end{abstract}

\maketitle

\section{The Chooz experimental site}
The experimental site is located in the Ardennes (France),
 close to the Chooz nuclear power plant, operated by the
French company Electricite de France (EDF). There are two N4 type PWR
 reactors of 4.27 GW$_{th}$ each. 
We will use two almost identical detectors,
containing a fiducial volume of 10 tons of liquid scintillator doped
with 0.1\% of Gadolinium (Gd). The laboratory of the first CHOOZ experiment,
located 1.05 km (the {\it Chooz-far} site, overburden of 300 m.w.e.) 
from the cores will be used again. This is the main
 advantage of this site. In order to cancel the systematic
errors originating from the nuclear reactors ($\overline{\nu_{e}}$
 flux and energy spectrum), 
as well as to reduce the systematic errors, a second detector will be installed
close to the nuclear cores (the \emph{Chooz-near} site). Since no
natural hills or underground cavity already exists at this location,
an artificial overburden of about 20 meters height has to be built. 
At 150~m the required overburden to protect the detector from cosmic ray induced
backgrounds is 60~m.w.e..
\section{The new detector concept}
\begin{center}
\begin{figure}[htb]
\vspace{9pt}
\begin{center}
\includegraphics[width=70mm]{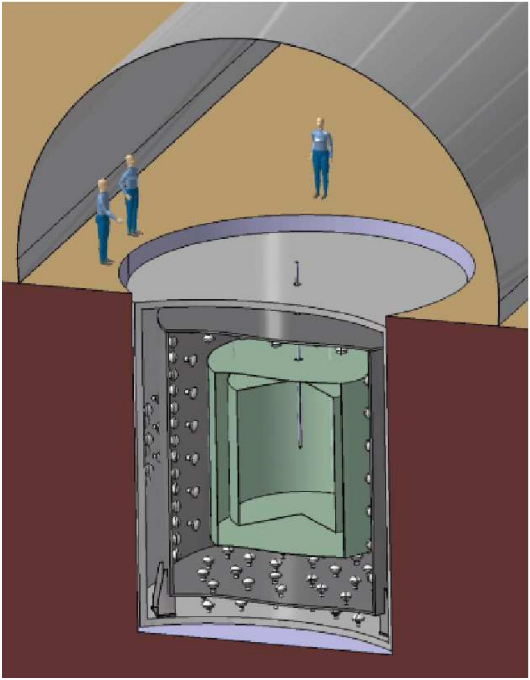}
\end{center}
\caption{The detector is located in the tank used
for the CHOOZ experiment (7 meters high and diameter). 
About 10 tons of a liquid scintillator doped with Gd is contained in a double-acrylic
cylinder surrounded by the gamma-catcher region, the buffer and the
muon veto. The optical coverage of the PMTs is about 15\%. }
\label{fig:detectorfar}
\end{figure}
\end{center}

The detector design is an evolution of the detector of the first
experiment \cite{choozlast}.
To improve the sensitivity of Double-Chooz with respect to CHOOZ it
is planned to increase statistics and to reduce and better control
the systematic errors and backgrounds. In order to increase the exposure
to 60,000 events at Chooz-far (statistical error of 0.4\%) it is planned
to use a target cylinder of 120 cm radius and 280 cm height, providing
a fiducial mass of 10 tons (12.7m$^{3}$), 2.3 times larger than in CHOOZ.
In addition, the data taking period will be extended to at least three
years, and the overall data taking efficiency will be improved. The
near and far detectors will be identical inside the PMT supporting
structure. This will allow a relative normalization systematic error
of 0.6\%. Starting from the center of the target the detector elements
are as follows (Figure~\ref{fig:detectorfar}).
The neutrino target: A 120 cm radius, 280 cm height, 8 mm thick acrylic
cylinder, filled with 0.1\% Gd loaded liquid scintillator.
The baseline of the scintillator being developed for the new experiment
is a mixture of 20\% of PXE and 80\% of dodecane, with small quantities
of PPO and bis-MSB added as fluors. 
The $\gamma$-catcher: A 60 cm buffer of liquid scintillator not loaded
with Gd, with the same light yield as the target. 
The role of this {\it new} region is to get the full positron
energy, as well as most of the neutron energy released after neutron
capture. It is enclosed in a 180 cm radius, 400 cm height, 10 mm thick
acrylic cylinder.
The non scintillating buffer: A 95 cm buffer of non scintillating
oil, to decrease the level of accidental background (mainly the
contribution from photomultiplier tubes radioactivity from potassium,
uranium and thorium) and the PMT supporting structure.
The outer veto: A 60 cm veto region filled with liquid scintillator
for the far detector, and a slightly larger one (about 100 cm) for
the near detector.
The external shielding: A 15 cm steel shielding surrounding the far
detector, and a $\sim$1 m low radioactive sand or water layer for
the near detector. 
We plan to build the double acrylic vessels at the manufacturer and
transport it to the detector sites in a single piece. This integration
procedure allow us to minimize the differences between the acrylic
vessels as well as to reduce the residual mechanical stress that could
favor the acrylic crazing.
\section{Systematic errors and backgrounds}
In the first CHOOZ experiment, the total systematic error amounted
to 2.7\%. Thanks to the use of the double detector concept, each error
originating from the neutrino source, e.g. the reactors, cancels.
Thus we can neglect the ``reactor cross section error'' of 1.9\%,
the uncertainty on the reactor power of 0.7\%, as well as the lack
of knowledge of the energy released per fission of 0.6\%. The dominant error
for Double-Chooz will thus be the relative normalization between the
two detectors; it originates for instance from the detection efficiency,
or a difference in the number of free protons contained in the acrylic
targets. We focus our efforts towards the precise measurement of the relative volumes between the 
acrylic targets, and the dead time measurement. In addition, the position
of the near detector with respect to the core will have to be measured
with a precision better than 10~cm. 
In Double-Chooz, we estimate the total systematic error on the normalization
between the detectors to be less than 0.6\%. The main contributions
come from the solid angle (0.2\%), the volume (0.2\%), the density and
H/C ratio (0.15\%), the neutron detection efficiency and energy measurement
(0.2\% and 0.1\%), the e$^{+}$-n time delay (0.1\%) and
the dead time (0.25\%, with a new method of fake triggers generated inside
the target under development).

The signature for a neutrino event is a prompt signal with a minimal
energy of about 1~MeV and a delayed 8~MeV signal after neutron capture
by a Gd nucleus. This may be mimicked by background events
which can be divided into two classes: accidental and correlated events.
The former can be reduced by a careful selection of the materials
used to build the detector. In addition this background is easy to measure
in-situ, and its subtraction lead to a small systemactic error. 
A comprehensive Monte-Carlo study shows that the correlated events are 
the most severe background source for the experiment. 
Our simulation reproduced fairly well the correlated 
background rate measured in the first CHOOZ experiment and is thus reliable.
Two processes mainly contribute: $\beta$-neutron cascades and very
fast external neutrons. Both types of events are coming from spallation
processes of high energy muons. In total the background rates for
the near detector will be between 9/d and 23/d, for 60
m.w.e. overburden. For the far detector a total background
rate between 1/d and 2/d can be estimated. This can be compared with
the signal of $\sim4000$/d and 80/d in the near and far detectors. 
The overburden of the near detector has been chosen in order to keep the signal to
background ratio above 100. Under this condition, even a knowledge
of the backgrounds within a factor two keeps the associated systematic
error below the percent. 
\section{Discovery potential}
The Double-Chooz experiment is searching for a deficit in
the $\overline{\nu_{e}}$ flux at 1.05 km from the cores,
 while the near detector monitors the $\overline{\nu_{e}}$ flux and 
energy spectrum prior to any neutrino oscillation. This disappearance
channel allows a ``clean'' measurement of $\sin(2\theta_{13})^{2}$. Assuming
a relative normalization error of 0.6\%, and three years of data taking
with typical live time for both reactor and detector operations, the
sensitivity will be $\sin(2\theta_{13})^{2} < 0.025$ (at 90$\%$ C.L.,
for $\Delta m_{atm}^{2}=2.4$ eV$^{2}$) in the case of no-oscillation.
The discovery potential with a so-called ``3-$\sigma$'' effect
will be around 0.04. For a true value $\sin(2\theta_{13})^{2}$=0.1
a rate only analysis will reject the no-oscillation scenario
at 2.6$\sigma$, whereas a shape+rate analysis will reject the no-oscillation
scenario at about 6$\sigma.$ This clearly shows that the shape distortion
could be used as a smoking gun in Double-Chooz. For a true value
$\sin(2\theta_{13})^{2}$=0.1, Double-Chooz would
perform a measurement of the oscillation parameter with a 67\%
relative error (90\% C.L.); this can be compared with the  
 potential of 100-130\% of the  complementary superbeam experiments.
\section{Double-Chooz and non-proliferation}
Recently, the International Atomic Energy Agency (IAEA) expressed
the interest to use anti-neutrinos as a tool to verify the non-proliferation
of nuclear weapons. A detector close to a nuclear plant, like Double-Chooz-near,
could provide ``remote'' and non-intrusive measurements of plutonium
content in reactors, since the antineutrino flux and energy spectrum
depend upon the thermal power and the fissile isotopic composition
of the reactor fuel. A part of the Double-Chooz collaboration is planing
new measurements of the spectrum of various fissile elements in order
to reduce the error of the neutrino spectrum emitted by a nuclear
core. For that purpose, a new $\beta-$spectrometer is being studied to
upgrade the Mini-Inca instrument, at the ILL research reactor in
Grenoble (France). 
Furthermore, the upgrade of the monitoring system of
the thermal power is under study; this could allow a better
monitoring of the $\overline{\nu_{e}}$ flux, and to measure the
position of the source barycenter within a few centimeters.
\section{Outlooks}
The Double-Chooz collaboration is composed of about 16 institutes
(in France, Germany, Italy, Russia and USA). A Letter of Intent has
been released in may 2004 \cite{DoubleChoozLOI}, and the experiment
has been approved in France. The funding of the near
laboratory in partnership with the EDF power company and the local
authorities is currently being discussed. If fully approved in
2005, it is intended to start taking data at Chooz-far in 2007, and
at Chooz-near in 2008. If the collaboration meets this goal,
Double-Chooz could provide a sensitivity limit of $\sin(2\theta_{13})^{2}<0.05$
(at 90$\%$ C.L.) within the year 2009, and 0.02-0.03 in 2011.


\begin{thebibliography}{9}
\bibitem{choozlast}M. Apollonio, \emph{et. al} (CHOOZ Collaboration), Eur. Phys. J. C27,
331 (2003). 
\bibitem{DoubleChoozLOI}F. Ardellier, \emph{et. al} (Double-Chooz Collaboration), hep-exp/0405032 (2004).
\end{thebibliography}
\end{document}